# Direct observation of electronic inhomogeneities induced by point defect disorder in manganite films.

**Short title: Observation of disorder induced "phase separation" in manganite films.**

M. Sirena[1], A. Zimmers[3], N. Haberkorn[1], E. Kaul[1], L. B. Steren[2], J. Lesueur[3], T. Wolf[3], Y. Le Gall[4] and J.-J. Grob[4] and G. Faini[5].

[1] Instituto Balseiro – Univ. Nac. de Cuyo & CNEA, Av. Bustillo 9510, 8400 Bariloche, Rio Negro – Argentina.

[2] Centro Atómico Constituyentes, Av. Gral. Paz 1499, San Martín 1650, Buenos Aires-Argentina.

[3] UPR5-LPEM-CNRS, Physique Quantique, E.S.P.C.I., 10 Rue Vauquelin, 75231 Paris, France.

[4] Institut d'Électronique du Solide et des Systèmes, UMR 7163, 23 rue du Loess – BP20, F-67037 Strasbourg Cedex 02, France.

[5] LPN-CNRS, Route de Nozay, 91460 Marcoussis, France

Abstract

We have investigated the influence of point defect disorder in the electronic properties of manganite films. Real-time mapping of ion irradiated samples conductivity was performed though conductive atomic force microscopy (CAFM). CAFM images show electronic inhomogeneities in the samples with different physical properties due to spatial fluctuations in the point defect distribution. As disorder increases, the distance between conducting regions increases and the metal-insulator transition shifts to lower temperatures. Transport properties in these systems can be interpreted in terms of a percolative model. The samples saturation magnetization decreases as the irradiation dose increases whereas the Curie temperature remains unchanged.

**Pacs:** 71.30.+h, 75.47.Lx

Since their rediscovery in the past decade [1], $A_{1-x}A'_{x}MnO3$ (A: La, Pr; A': Sr, Ba, Ca) manganites have attracted a lot of attention in the scientific community. The renewed interest in these materials is due to their colossal magnetoresistance (CMR) effect and its possible technological applications. From early works it became clear that disorder, through changes in the Mn-O distance and the Mn-O-Mn angle, plays an important role in the transport and magnetic properties of bulk and thin manganite films [2-8]. Many results on this subject were explained in terms of phase-separation or phase coexistence [9-11] in these systems. However, the origin of this phenomenon is still a matter of discussion. Shenoy et. al. recently proposed that Coulomb interactions may be responsible of electronic inhomogeneities in manganites rather than disorder or phase competition [12]. Conductive atomic force microscopy (CAFM) [13], through direct mapping of the system's conductivity has proven to be a powerful tool to study this kind of problem. Last year, phase separation in polycrystalline manganite films was established by this measurement method [14].

In this work, vacancy-interstitial pairs were introduced by ion implantation into ferromagnetic $La_{0.75}Sr_{0.25}MnO3$ thin films in order to induce lattice disorder in a controlled way [15,16]. Ion irradiation has recover importance for the developing of new technological devices like planar tunnel junctions. Typically ferromagnetic tunnel junctions present a current perpendicular to the plane configuration. However, their fabrication is complex, involves several lithography steps, and suffers from pinholes in the thin insulating barrier. Irradiation has been successfully used by the authors to fabricate planar HTc Josephson Junctions and high Tc supeconducting devices [17] which present the same technological challenges than oxides tunnel junctions, from the fabrication point of view. The irradiation can be used to render insulating the manganite and draw the mesa circuits. Planar tunnel junctions can be fabricated creating a thin insulator barrier by ion irradiation through a photoresist mask with a small aperture across done with electronic lithography. The planar tunnel junctions can be placed directly and easily in the circuit opening the corresponding holes in the photoresist mask. Of course, studying the influence of ion irradiation in the physical properties of manganite films becomes very important to achieve this objective. CAFM was used to obtain direct images of the influence of point defect disorder in the phase-separation of manganite films. Transport and magnetic properties of the samples were analyzed in the frame of this model.

Thin 50nm $La_{0.75}Sr_{0.25}MnO3$ films were grown on single-crystal $SrTiO_3$ (100) substrates by DC magnetron sputtering from a stoichiometric ceramic target. More details about the films' fabrication method and characterization are given elsewhere [18]. The films were irradiated with oxygen ions ($O^+$) at 150 keV with different doses, ranging from $\phi= 1 \times 10^{14}$ ion.cm$^2$ to $\phi= 10 \times 10^{14}$ ion.cm$^2$. The choice of the oxygen energy was made considering the authors previous experience with oxygen implantation into HTc materials. The 150keV O+ ions are not expected to alter the material, either chemically or physically. The ions can be stopped by metallic or photoresist masks of appropriate thickness, which is important as mentioned before for the fabrication of different devices [17]. The damage profile calculated by TRIM simulations [19] is constant along the film depth for 50nm films thick. The resistivity of the samples has been measured using a standard four-probes configuration. The temperature and field dependence of the magnetization have been studied using a Superconductor Quantum Interference Device (SQUID) magnetometer. CAFM measurements were performed in a Veeco Dimension 3100 ® SPM with a CAFM module. The scans were performed using a diamond doped with boron conductive tip in contact mode. The measurements have been made keeping the probe polarization voltage fixed to 0.5 V, with the same deflection setpoint (0.4V). Tests, performed with different probe polarizations and deflection setpoints have been done in order to assure that CAFM results were independent of measurements conditions.

Figure 1 presents the topographic (left) and CAFM current (right) images of irradiated LSMO films. The pristine samples present very low roughness (~0.2 nm). The images put in evidence the correlation between the irradiation of the samples and their roughness: the increase of the irradiation dose leads to an augmentation of the films roughness reaching 4 nm for the highest dose ($\phi= 10 \times 10^{14}$ ion.cm$^2$). CAFM images show that even the non-irradiated sample presents a finite and spatially inhomogeneous distribution of conductivities. We define conducting regions as those that show non-zero CAFM current while insulator zones represent those where there no CAFM current is observed for the standard measuring conditions. In the images, the conducting islands are clear while the insulator regions are dark. The scale lengths of the conducting and insulator regions in the pristine sample are clearly different from the ones measured in the irradiated samples, even for the lowest irradiation dose. The origin of the inhomegeneities in pristine thin films may associated to the intrinsic *A* cation disorder and to inhomogeneous strains fields due to the different A cation radii in manganite compounds. On the other hand, the electronic inhomogeneity observed in irradiated samples would be probably associated to the random distribution of point defect, strain fields induced by the ion bombardment or inhomogeneous changes of the Mn$^{3+/4+}$ ratio to reduce lattice strains in the irradiated areas.

The influence of point defect disorder in the conductivity of the irradiated samples can be clearly seen in a histogram of conducting surface as function of the CAFM current, i.e. conductivity (shown in Figure 2 for three irradiation doses). The histograms present a single continuum distribution of conductivities. Continuum conductivity histograms were also found by STM measurements on Pr$_{0.68}$Pb$_{0.32}$MnO$_3$ single crystals [20] and seem to indicate that there is not a phase separation in the system. Instead, there would be an arrangement of regions with proper characteristics and physical properties due to the inhomogeneous distribution of strains and point defects at the nanoscale. I(V) (Figure 3) curves for the so-called conducting islands and insulator regions are non linear in agreement with recent results obtained on polycrystalline films [14]. However, it is clearly discernible from these curves that the energy gap in the conducting regions (Figure 3_a) is much smaller than the one measured for the insulator regions (Figure 3_b), when the irradiation dose is increases the gap also increases (figure 3_c). At room temperature, the carries in the different regions are still localized but the energy "gap" is smaller for reduced disorder. This mean for instance that for the same applied voltage the CAFM current is higher in the "metallic" regions.

The pristine sample is ferromagnetic with a transition temperature, Tc of 320 K and a total saturation magnetization of 545 emu/cm$^3$ in agreement with the calculated value considering the $Mn^{3+}/Mn^{4+}$ ratio and its lattice parameter. The sample presents a metal-insulator transition at a characteristic temperature ($T_{MI}$) around the Tc, as expected from the double exchange model [21]. As the irradiation dose increases the saturation magnetization decreases but the Tc doesn't change. However, the $T_{MI}$ systematically decreases for increasing irradiation doses and for the highest dose the sample becomes insulator in all the measured temperature range. This $T_{MI}$ decoupling from the magnetic transition temperature is in contradiction with the early believe that for ion irradiated films $T_C$ follows the $T_{MI}$ [15], but has already been observed in previous works as a result of increasing disorder [22]. A detailed study of the magnetic and transport properties of the ion irradiated films will be presented shortly [23]. Figure 4 presents the saturation magnetization, measured at 5K and the magnetic transition temperature for the samples with different irradiation doses, as function of the total conducting area obtained from the CAFM images. The total non-conducting area is obtain integrating the surface of the areas presenting non CAFM current. A clear correlation between the saturation magnetization and the total conducting area can be observed: as the irradiation dose increases, the total conducting area decreases and there is a reduction of the saturation magnetization. This suggests that the more conducting regions are related to the ferromagnetism in the sample. In the same way that not only the size of these regions but also their conductivity decreases with increasing irradiation dose, it is probable that the decrease of the films magnetizations is related not only to the reduction of the islands size but also to a reduction of the microscopic magnetization within the conducting regions. This gives place to an exponential decrease of the magnetization instead of a linear decrease as expected considering only the size reduction of these areas.

Figure 5 shows the metal-insulator transition temperature obtained from resistance versus temperature measurements as function of the mean distance between conducting regions obtained graphically from the CAFM images. We have measured with the software the border to border distance of around 30 conducting areas of different sizes and distances in order to estimate its mean value and dispersion. As the irradiation dose increases the distance between more conducting areas increases slowly. Figure 3 indicates that as the distance between the conducting islands increases the films $T_{MI}$ decreases. The metal-insulator transition in these inhomogeneous systems can be described with a percolative model [24,25]. When the temperature decreases the insulating region gradually increases due to the decrease of the thermally activated conductivity. However, below a critical temperature these areas increase their conductivity and their size [14]. At the macroscopic $T_{MI}$ a percolation of the conducting islands occurs. As the point defect disorder increases, lower temperatures are required for the system to undergo the percolative transition. This explains the decoupling between the magnetic transition and the metal-insulator transition observed in these systems. Generally, ion irradiation creates vacancy-interstitial (cation and oxygen) pairs, locally inducing lattice strain enhancement [26] and $Mn^{3+}/Mn^{4+}$ ratio depletion [27] which are responsible for the observed results.

We have found that point defect disorder influences in a complex way the transport and magnetic properties of manganites films. Increasing the density of vacancy-interstitial pairs not only increases the systems resistance but also increases its electronic inhomogeneity. Transport properties in these systems can be interpreted in terms of a percolative model, as disorder increases the distance between conducting regions increases, decreasing the macroscopic metal-insulation transition temperature. We have also found that as disorder increases, the saturation magnetization of the samples decreases. However, the magnetic transition temperature remains unchanged.

The authors acknowledge K. Bouzehouane and S. Fusil for the formation received in CAFM measurements; Micra and Veeco crew for extraordinary technical support. This work was partially supported by the ANPCYT (PICT 06 2092, PICT 05 33304). M. S, N. H, L.B.S. are members of CONICET, Argentina. L.B.S is a fellow of the Guggenheim Foundation.

**References.**

[1] R.V. Helmholt et al., Phys. Rev. Lett. **71** (1993) 2331

[2] J. Goodenough, *Interscience Monographs on Chemistry* (Wiley, New York, 1963), Vol. I.

[3] A. Millis, T. Darling, A. Migliori, J. Appl. Phys., **83**, (1998) 1588 .

[4] J. Fontcuberta, V. Laukhin, and X. Obradors, Appl. Phys. Lett., **72**, (1998) 2607.

[5] L. Rodriguez-Martinez and J. P. Attfield, Phys. Rev. B **54**, (1996) R15622.

[6] W. Zhang, X. Wang, M. Elliott, and I. Boyd, Phys. Rev. B **58**, (1998) 14143.

[7] H. Ju, K. Krishnan, and D. Lederman, J. Appl. Phys. **83**, (1998) 7073.

[8] J. Aarts, S. Freisen, R. Hendrikx, H. Zandbergen, Appl. Phys. Lett. **72,** (1998) 2975.

[9] T. Becker, C. Streng, Y. Luo, V. Moshnyaga, B. Damaschke, N. Shannon and K. Samwer, Phys. Rev. Lett. **89**, (2002) 237203.

[10] Amlan Biswas, M. Rajeswari, R. C. Srivastava, T. Venkatesan, R. L. Greene, Q. Lu, A. L. de Lozanne and J. Millis, Phys. Rev. B. **63**, (2001) 184424.

[11] A. Tebano, C. Aruta, P. G. Medaglia, F. Tozzi, G. Balestrino, A. A. Sidorenko, G. Alloni, R. De Renzi, G. Ghiringhelli, C. Dallera and L. Baricovich, Phys. Rev. B. **74**, (2006) 245116.

[12] V. B. Shenoy, T. Gupta, H. R. Krishnamurthy and T. V. Ramakrishnam, Phys. Rev. Lett., **98**, (2007) 097201.

[13] M. Bibes, M. Bowen, A. Barthelemy, A. Anane, K. Bouzehouane, C. Carretero, E. Jacket, J. P. Contour and O. Durand, Appl. Phys. Lett., **82**, (2003) 3269.

[14] Y. H. Chen and T. B. Wu, Appl. Phys. Lett. **93**, (2008) 224104.

[15] C.-H Chen, V. Talyansky, C. Kwon, M. Rajeswari, R. P. Sharma, R. Ramesh, T. Venkatesan, J. Meingailis, Z. Zhang and W. Chu, Appl. Phys. Lett. **69**, (1996) 3089.

[16] J. F. Gibbons, Proc. IEEE **60,** (1972) 1062.

[17] N. Bergeal, J. Lesueur, M. Sirena, G. Faini, M. Aprili, J. P. Contour and B. Leridon, J. Appl. Phys. **102**, (2007) 083903.

[18] M. Sirena, L. Steren and J. Guimpel, Thin Solid Films, **373**, (2000) 102.

[19] J. F. Ziegler and J. P. Biersack, J. Exp. Theor. Phys. **87**, (1998) 375

[20] S. Rößler, S. Ernst, B. Padmanabhan, Suja. Elizabeth, H. L. Bhat, F. Steglich and S. Wirth, Europhysics Letters, **83**, (2008) 17009

[21] C. Zener, Phys. Rev. 81, (1951) 440.

[22] L. B. Steren, M. Sirena, J. Guimpel, J. Magn. Magn. Mater., **211,** (2000) 28.

[23] M. Sirena, A. Zimmers, N. Haberkorn, E. Kaul, L. B. Steren, J. Lesueur, T. Wolf, Y. Le Gall and  J.-J. Grob and G. Faini; inpreparation.

[24] M. Uehara, S. Mori, C. H. Chen and S. W. Cheong, Nature **399**, (1999) 560.

[25] N. Mathur and P. Littlewood, Phys. Today, **56**, (2003) 25.

[26] R. Bathe, K. P. Adhi, S. I. Patil, G. Marest, B. Honneyer and S. Ogale, Appl. Phys. Lett. **76**, (2000) 2104.

[27] R. Bathe, S. I. Patil, K. P. Adhi, B. Hannoyer and G. Marest, J. Appl. Phys. **93**, (2003) 1127.

**Figure 1:** Topographic (left) and CAFM (right) images (5 μm x 5 μm) of irradiated LSMO films for different irradiation doses: (a) ϕ=0 ion/cm$^2$, (b) ϕ= 3x10$^{14}$ ion/cm$^2$, (c) ϕ= 7.5x10$^{14}$ ion/cm$^2$ and (d) ϕ= 10x10$^{14}$ ion/cm$^2$.

**Figure 2:** Histograms of the covered surface for different values of the CAFM current (i.e. conductivity) of the irradiated LSMO samples at room temperature.

**Figure 3:** Typical I(V) curves for (a) the conducting regions corresponding to the sample irradiated with a dose of 3x10$^{14}$ ion/cm$^2$ , (b) insulating regions of the same sample and (c) insulating regions of the sample irradiated with 4.5x10$^{14}$ ion/cm$^2$ (c).

**Figure 4:** Saturation magnetization and magnetic transition temperature as function of the covered conducting surface at room temperature. The inset shows a typical detail of the CAFM image with highlighted conducting and insulating regions. Lines are the best linear fit of the experimental data.

**Figure 5:** Metal-insulator transition temperature as function of the mean distance between conducting regions. The line is a linear fit of the data.

Figure 1: Sirena et. al.

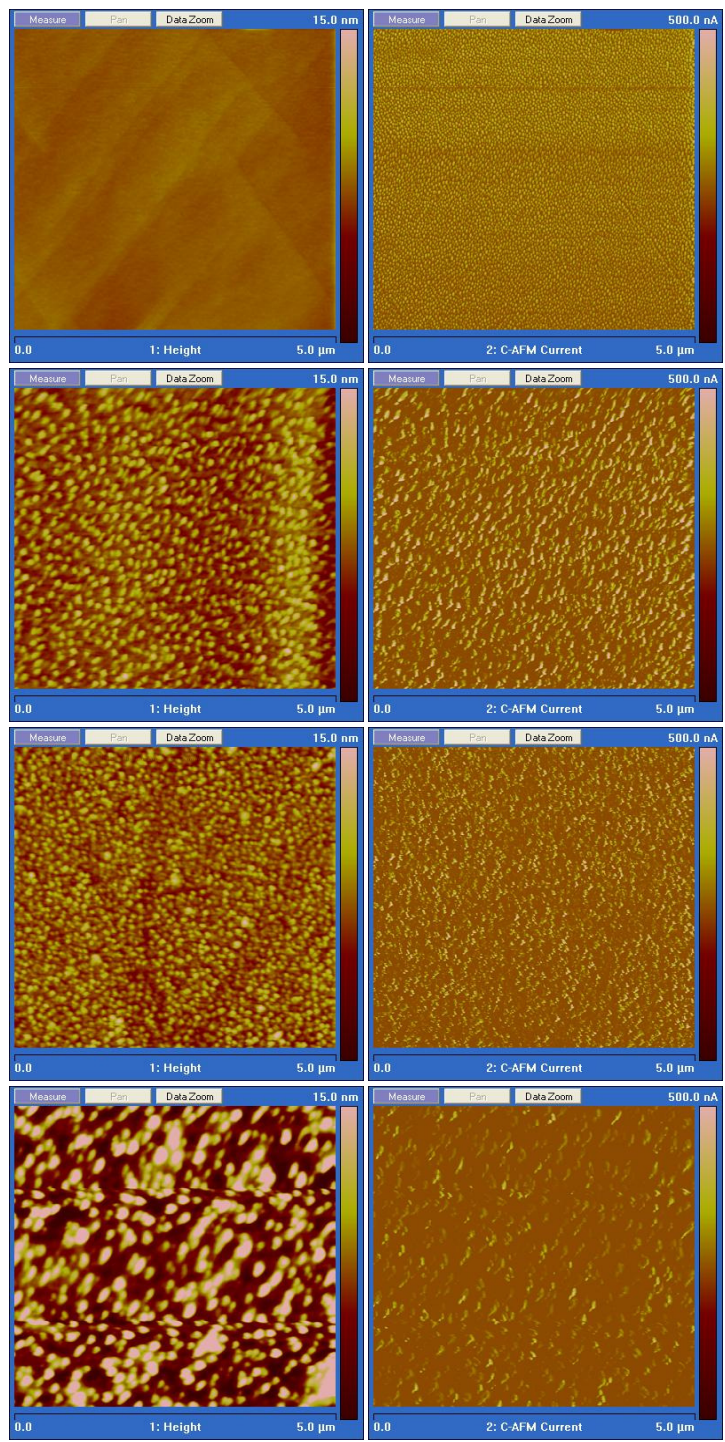

Figure 2 : Sirena et. al.

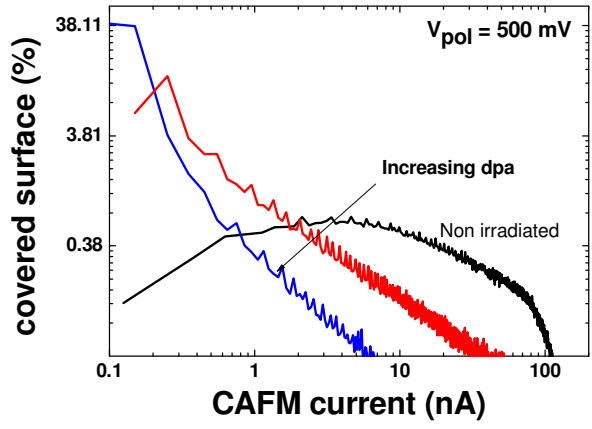

Figure 3: sirena et. al.

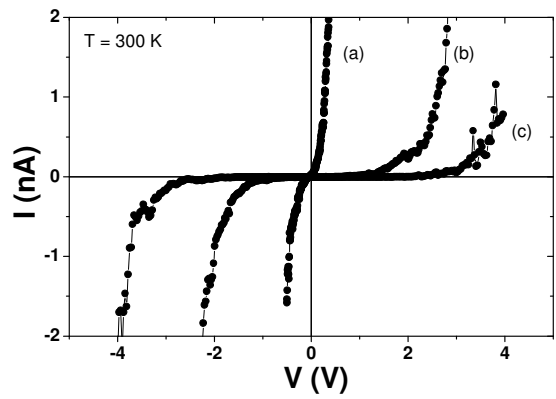

Figure 4: Sirena et. al.

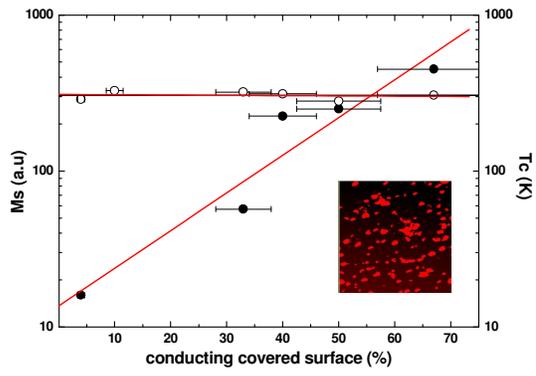

Figure 5: Sirena et. al.

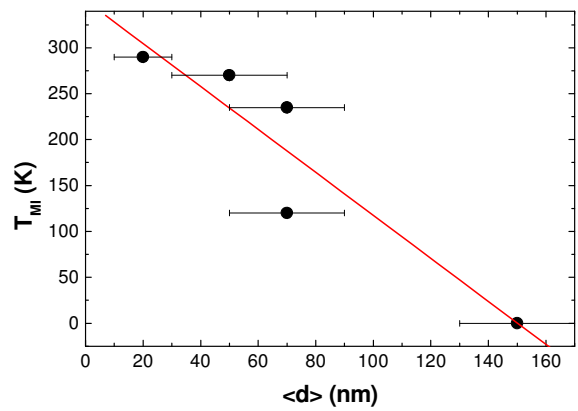